\documentclass[12pt,preprint]{aastex}

\newcommand{\sgr}{\mbox{SGR\,J1550--5418}}
\newcommand{\psr}{\mbox{PSR\,J1550--5418}}
\newcommand{\esrc}{\mbox{1E\,1547.0--5408}}
\newcommand{\trig}{$T_0$}
\newcommand{\degrees}{\ensuremath{^\circ}}

\begin{document}

\title{Magnetar Twists: {\it Fermi}/Gamma-ray Burst Monitor (GBM) detection of  \sgr}

\author{Yuki~Kaneko\altaffilmark{1},
         Ersin~G\"o\u{g}\"u\c{s}\altaffilmark{1},
         Chryssa~Kouveliotou\altaffilmark{2},
 	Jonathan~Granot\altaffilmark{3},
         Enrico~Ramirez-Ruiz\altaffilmark{4},
         Alexander~J.~van~der~Horst\altaffilmark{2},
         Anna L. Watts\altaffilmark{5},
         Mark H. Finger\altaffilmark{6},
         Neil Gehrels\altaffilmark{7},
         Asaf Pe'er\altaffilmark{8},
         Michiel van der Klis\altaffilmark{5},
         Andreas von Kienlin\altaffilmark{9},
         Stefanie Wachter\altaffilmark{10},
         Colleen A. Wilson-Hodge\altaffilmark{2},
         Peter M. Woods\altaffilmark{11}
         }

\altaffiltext{1}{Sabanc\i~University, Orhanl\i-Tuzla, \.Istanbul 34956, Turkey}
\altaffiltext{2}{Space Science Office, VP62, NASA/Marshall Space Flight Center, Huntsville, AL 35812, USA}
\altaffiltext{3}{Centre for Astrophysics Research, University of Hertfordshire, College Lane, Hatfield AL10 9AB, UK}
\altaffiltext{4}{Department of Astronomy and Astrophysics, University of California, Santa Cruz, CA 95064, USA}
\altaffiltext{5}{Astronomical Institute ÒAnton Pannekoek,Ó University of Amsterdam, PO Box 94249, 1090 GE Amsterdam, The Netherlands}
\altaffiltext{6}{Universities Space Research Associations, NSSTC, Huntsville, AL 35805, USA}
\altaffiltext{7}{NASA/Goddard Space Flight Center, Greenbelt, MD 20771, USA}
\altaffiltext{8}{Space Telescope Science Institute, 3700 San Martin Dr., Baltimore, MD, 21218, USA}
\altaffiltext{9}{Max-Planck Institut f\"ur extraterrestrische Physik, 85748 Garching, Germany}
\altaffiltext{10}{Spitzer Science Center/California Institute of Technology, Pasadena, CA 91125, USA}
\altaffiltext{11}{Dynetics, Inc., 1000 Explorer Boulevard, Huntsville, AL 35806, USA}
\email{yuki@sabanciuniv.edu}

%%%%%%%%%%%%%%%%%%%%%%%%%%%%%%%%%%%%%%%%%%%%%%%%%%%%%%%%%%%%%%%%%%%%%%%%%%%%%%%
%%%%%%%%%%%%%%%%%%%%%%%%%%%%%%%%%%% Abstract %%%%%%%%%%%%%%%%%%%%%%%%%%%%%%%%%% 
%%%%%%%%%%%%%%%%%%%%%%%%%%%%%%%%%%%%%%%%%%%%%%%%%%%%%%%%%%%%%%%%%%%%%%%%%%%%%%%
\begin{abstract}

\sgr~(previously known as AXP\,\esrc~or \psr) went into three active 
bursting episodes in 2008 October and in 2009 January and March,
emitting hundreds of typical Soft Gamma Repeater (SGR) bursts in soft
gamma-rays.  The second episode was especially intense, and our
untriggered burst search on {\it Fermi}/GBM data (8$-$1000~keV)
revealed $\sim$450 bursts emitted over 24 hours during the peak of
this activity. Using the GBM data, we identified a $\sim$150-s-long
enhanced persistent emission during 2009 January 22 that exhibited
intriguing timing and spectral properties: (i) clear pulsations up to
$\sim$110~keV at the spin period of the neutron star ($P \sim 2.07$~s,
the fastest of all magnetars), (ii) an additional (to a power-law)
blackbody component required for the enhanced emission spectra with
$kT \sim 17$~keV, (iii) pulsed fraction that is strongly energy
dependent and highest in the 50$-$74~keV energy band.  A total
isotropic-equivalent energy emitted during this enhanced emission is
estimated to be $2.9 \times 10^{40}(D/5\,{\rm kpc})^2$~erg.  
%We conclude that the enhanced emission detected in the persistent flux of \sgr~may be a transitional event between an intermediate SGR event and
%a giant flare. 
The estimated area of the blackbody emitting
region of $\approx 0.046(D/5\,{\rm kpc})^2\;{\rm km}^2$ (roughly a few
$\times 10^{-5}$ of the neutron star area) is the smallest ``hot spot"
ever measured for a magnetar and most likely corresponds to the size
of magnetically-confined plasma near the neutron star surface.

\end{abstract}


%%%%%%%%%%%%%%%%%%%%%%%%%%%%%%%%%%%%%%%%%%%%%%%%%%%%%%%%%%%%%%%%%%%%%%%%%%%%%%%
%%%%%%%%%%%%%%%%%%%%%%%%%%%%%%%%%%%% INTRO %%%%%%%%%%%%%%%%%%%%%%%%%%%%%%%%%%%% 
%%%%%%%%%%%%%%%%%%%%%%%%%%%%%%%%%%%%%%%%%%%%%%%%%%%%%%%%%%%%%%%%%%%%%%%%%%%%%%%
\section{Introduction}

A very small group (roughly half a dozen) of isolated neutron stars
have manifested themselves in one class as Soft Gamma Repeaters (SGRs)
linked by numerous common distinguishing properties. Among the most
characteristic SGR attributes are (i) X-ray luminosities much larger
(by $\sim$100 times) than the ones expected from their rotational
energy losses, and (ii) the emission of repeated bursts of soft gamma
rays. SGR bursts range from ``typical'' short events lasting
$\sim$0.1~s with peak luminosities of $L_p
\lesssim10^{41}$~erg~s$^{-1}$, to occasional intermediate flares
lasting a few seconds with $L_p \sim10^{42}$$-$$10^{43}$~erg~s$^{-1}$,
and finally to -- extremely rare -- giant flares lasting a few hundred
seconds with $L_p \gtrsim10^{45}$~erg~s$^{-1}$.  SGRs were identified
together with Anomalous X-ray Pulsars (AXP) as ``magnetars'': neutron
stars powered by their extremely strong magnetic fields (surface
dipole $B \sim 10^{14-15}$~G; \citep{dun92, kou98}. Comprehensive
reviews on magnetars can be found in \citet{woo06} and \citet{mer08},
and references therein.

\esrc~was observed with the X-ray Multi-Mirror Mission (XMM-Newton) in 
2004 as a magnetar candidate, selected for its galactic plane location
and its relatively soft magnetar-like spectrum as seen with the
Advanced Satellite for Cosmology and Astrophysics (ASCA) during their
Galactic plane survey \citep{sug01}. Although no period was detected
in the original and follow-up XMM observations, \citet{gel07} also
proposed \esrc~as a magnetar candidate based on its spectrum and its
positional coincidence with an extended galactic radio source
G327.24-0.13 (possibly a supernova remnant). The subsequent discovery
in radio observations of a spin period of 2.07~s and a period
derivative of 2.3 $\times$~10$^{-11}$~s~s$^{-1}$ led to an estimated
dipole surface field of $B\sim2.2\times10^{14}$ G and confirmed the
source's magnetar nature; the source was also renamed as
\psr~\citep{cam07}. Its period makes \esrc~the fastest rotating
magnetar; the source is also one of the only two that emit in radio
wavelengths \citep[the other source is an AXP, XTE
J1810-197;][]{hal05, cam06}. The distance of the source has been
estimated by different authors using various methods: \citet{cam07}
found a distance of $\approx 9$~kpc by measuring radio dispersion;
\citet{gel07} estimated $\approx 4$~kpc assuming an association of the
source with a possible supernova remnant, G327.24-0.13; and most
recently \citet{tie09} reported an average distance of 4$-$5~kpc by
using observations of an X-ray scattering halo in the {\it Swift}/XRT
data. Throughout this paper we use $D_5 = D/5$~kpc as the source
distance measure.

On 2008 October 3, \esrc~entered an episode of X-ray activity,
emitting several typical SGR-like bursts over the next 7 days. During
this period, 22 short duration bursts were observed with the Gamma-ray
Burst Monitor (GBM) on board the {\it Fermi} Gamma-ray Space
Telescope. A detailed analysis of these events is presented in A. von
Kienlin~et~al. (2010, in preparation).

On 2009 January 22, the source entered a second period of extremely
high X-ray burst activity \citep{mer09}.  During the first 24~hours of
this ``storm", the {\it Fermi}/GBM triggered on the source 41 times:
the number of triggers was limited only by the instrument's capability
and did not reflect the actual number of bursts emitted by the
source. In fact, our on-ground search for untriggered events revealed
a total of $\sim$450 bursts during this 24 hour period: an unusually
high burst frequency from a single source (A.J. van der
Horst~et~al. 2010, in preparation). Based on this SGR-like behaviour,
we renamed the source as \sgr~\citep{kou09}.

Upon examination of the data from the first GBM trigger on January 22,
we identified 29 short events riding on an enhancement of the
underlying persistent emission lasting $\sim$150~s. Closer inspection
of this enhancement in different energy ranges revealed periodic
oscillations with a period consistent with the spin period of \sgr. We
present here a detailed temporal and spectral analysis of this
enhanced emission period. In \S\ref{sec:obs}, we briefly describe our
observations and the GBM instrument and data types. We present our
temporal analysis results in \S\ref{sec:tempo}, and our spectral
studies in \S\ref{sec:spec}. Finally we discuss the physical
implications of our discovery in \S\ref{sec:sum}.

%%%%%%%%%%%%%%%%%%%%%%%%%%%%%%%%%%%%%%%%%%%%%%%%%%%%%%%%%%%%%%%%%%%%%%%%%%%%%%%
%%%%%%%%%%%%%%%%%%%%%%%%%%%%%%%%%% SECTION 1 %%%%%%%%%%%%%%%%%%%%%%%%%%%%%%%%%% 
%%%%%%%%%%%%%%%%%%%%%%%%%%%%%%%%%%%%%%%%%%%%%%%%%%%%%%%%%%%%%%%%%%%%%%%%%%%%%%%
\section{Instrumentation and Data}\label{sec:obs}

The {\it Fermi}/GBM consists of 12 NaI detectors (8$-$1000~keV)
arranged in 4 clusters of three each and 2 BGO detectors (0.20$-$40~MeV) at opposite sides of the spacecraft \citep[for a detailed
description of the instrument, see][]{mee09}. GBM is currently the
only instrument with continuous broad-band energy coverage (8
keV$-$40~MeV) and a wide field of view (8~sr after taking into account
occultation by the Earth) and is, therefore, uniquely positioned to
accomplish a comprehensive magnetar (or any transient event)
monitoring. In trigger mode, GBM provides three types of data; CTIME
Burst, CSPEC Burst, and Time Tagged Event (TTE) data
\citep{mee09}. The CTIME Burst data have a time resolution of 64~ms
with rather coarse spectral information (8~energy channels). The CSPEC
Burst data provide high-resolution spectra (128~energy channels)
collected every 1.024~s. Both CTIME Burst and CSPEC Burst accumulate
data for $\sim$600~s after a trigger. The TTE data provide time-tagged
photon event lists for an accumulation time of 330~s, starting 30 s
prior to the trigger time; this data type provides a superior temporal
resolution down to 2$\mu$s at the same spectral resolution as the
CSPEC Burst data.

\begin{figure}[htbp]
%\epsscale{0.9}
\centerline{
\plotone{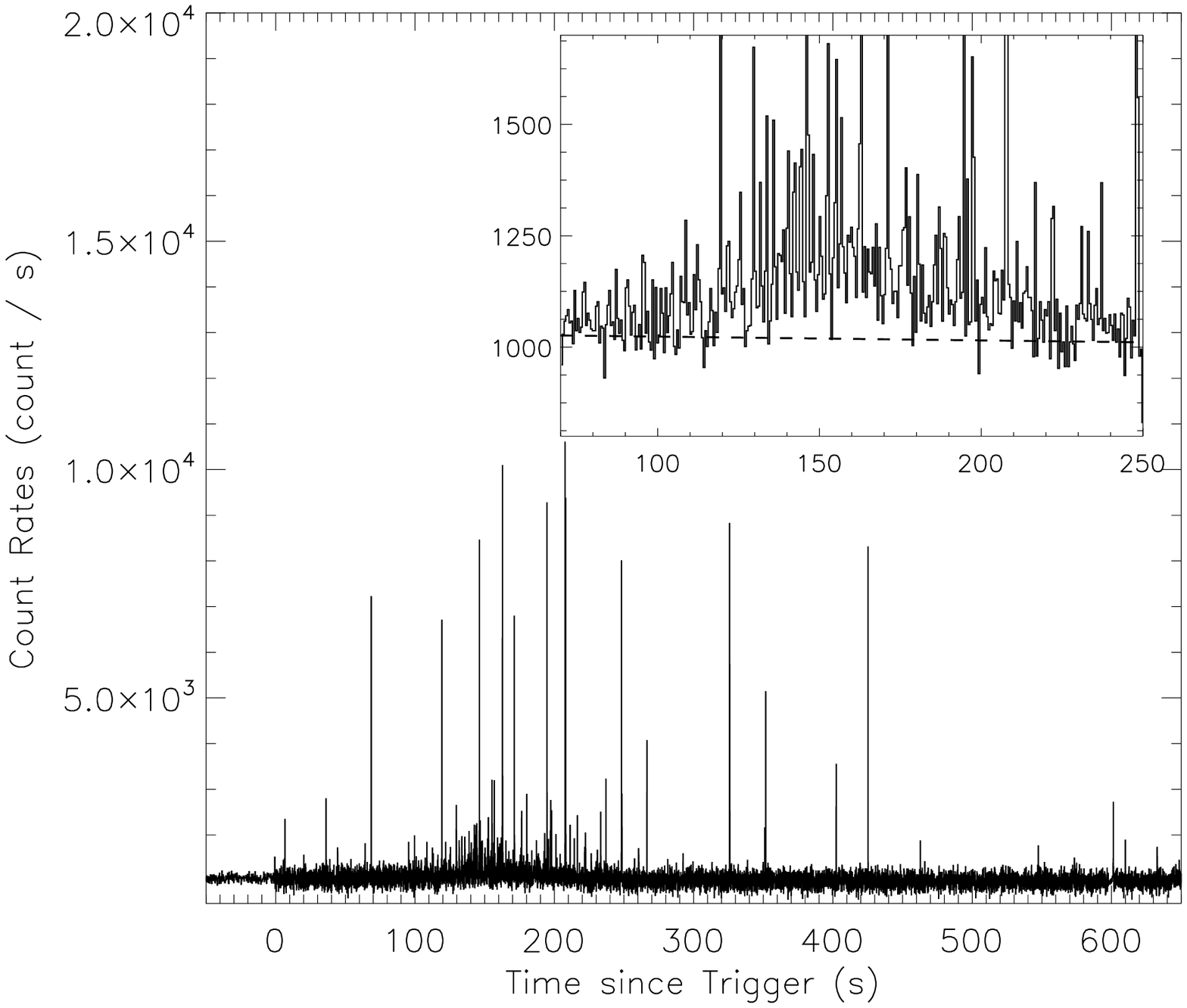}}
\caption{Lightcurve of \sgr~in 12$-$293~keV (GBM NaI\,0 CTIME data channels 1$-$4).  An enlarged view of the pulsed, enhanced emission is
shown in the inset.  The dashed line indicates the background level.}
\label{fig:lc_ch14}
\end{figure}

The first GBM trigger at the onset of the second active episode from
\sgr~was on 2009 January 22 at 00:53:52.17 UT (= \trig, GBM trigger
number 090122037). In the 600~s of the trigger readout we detected
many individual short bursts using our on-ground untriggered burst
search algorithm.  To accept an event as an untriggered burst, we
required excess count rates of at least 5.5$\sigma$ and 4.5$\sigma$ in
the first and second brightest detectors, respectively, in the 10$-$300~keV energy range. We used CTIME data in both continuous (256~ms time
resolution) and Burst mode (64~ms resolution).  Subsequently, we inspected energy-resolved burst morphology and compared each
detector zenith angle to the source for all 12 detectors, to determine
whether the events originated from \sgr. In total we identified about
a dozen very bright bursts and over 40 less intense bursts within
600~s after \trig~(see Figure \ref{fig:lc_ch14}). During the same
trigger readout we also discovered an enhancement in the underlying
persistent emission starting at approximately \trig$+ 70$~s and
lasting for $\sim$150~s (see inset of Figure \ref{fig:lc_ch14}).

One of the events recorded during these 600 s, specifically the burst
at \trig$+ 147$~s, was so bright that it initiated an Autonomous
Repoint Recommendation (ARR), causing the spacecraft to start slewing
towards the \sgr~direction. As the source was already close to the
boresight of the LAT, the slew angle was pretty small. However, we
proceeded to check whether the observed emission enhancement was
artificially caused by the spacecraft slewing.  First, we calculated
the variation in time of the zenith angle of \sgr~for each of the 12
GBM detectors.  At the onset of the enhancement (\trig$+ 70$~s), the
NaI\,0 detector had the smallest zenith angle to the source of
15\degrees.  Due to the ARR, NaI\,0 kept a constant angle of
18\degrees~to the source from \trig$+150$~s to \trig$+210$~s, after
which it constantly slewed away from the source until it reached an
angle of 23\degrees~at \trig$+ 270$~s.  During this time the
persistent emission kept rising until \trig$+ 150$~s, which alone
confirms that the enhanced emission is intrinsic to \sgr. The source
was in the field of view of the detector until $\sim$\trig$+ 225$~s,
at which time it went into an occultation by the LAT.  At the same
time NaI\,6 was slewing towards the source at an angle of 20\degrees,
but the emission was unfortunately obscured by the LAT until
$\sim$\trig$+225$~s.  We note, however, that the enhanced emission
was not clearly detected with NaI\,6 after \trig$+ 225$~s.  Based on the above, we conclude that the rise of
the enhanced emission was definitely not caused by the spacecraft slew
but by the source itself; we cannot unambiguously determine the end of
the decay trend (or the total duration of the enhancement) in the data
due to LAT obscuration.  In the
analysis presented in this paper, we have exclusively used data from
NaI\,0 (unless noted otherwise), to avoid any obscuration effects.
Finally, we also checked the LAT data (20~MeV$-$300~GeV) of the entire
day for associated high-energy gamma-ray emission, but found no
evidence of high-energy photons originating from the direction of
\sgr.

%%%%%%%%%%%%%%%%%%%%%%%%%%%%%%%%%%%%%%%%%%%%%%%%%%%%%%%%%%%%%%%%%%%%%%%%%%%%%%%
%%%%%%%%%%%%%%%%%%%%%%%%%%%%%%%%%% SECTION 2 %%%%%%%%%%%%%%%%%%%%%%%%%%%%%%%%%% 
%%%%%%%%%%%%%%%%%%%%%%%%%%%%%%%%%%%%%%%%%%%%%%%%%%%%%%%%%%%%%%%%%%%%%%%%%%%%%%%
\section{Temporal Properties of Pulsed Hard X-rays} \label{sec:tempo}

\subsection{Timing Analysis}

\begin{figure}[htbp]
%\epsscale{0.8}
\centerline{
\plotone{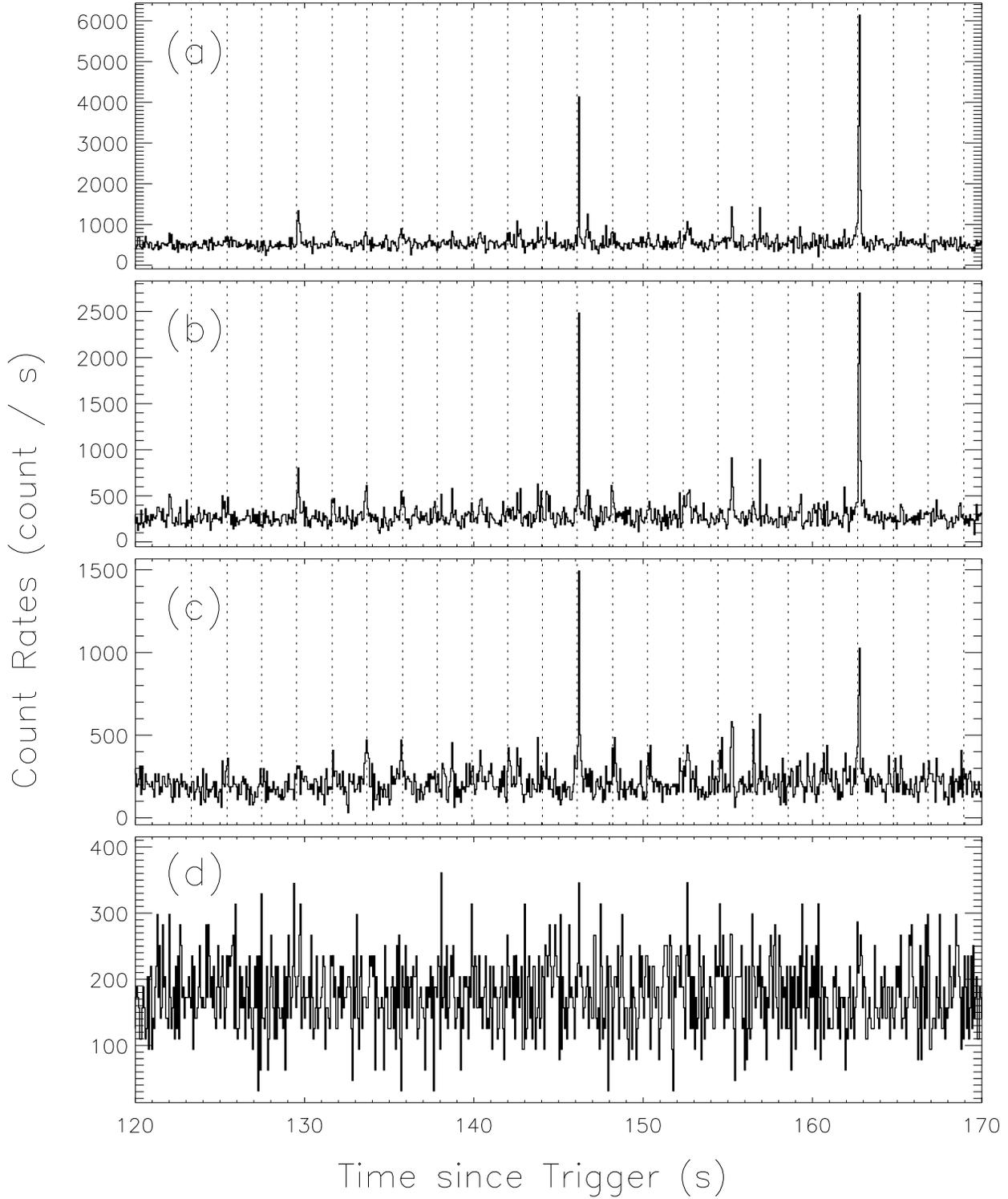}}
\caption{Lightcurve of \sgr~in various energy ranges;
(a) 12$-$27~keV, (b) 27$-$50~keV, (c) 50$-$102~keV, and (d)
102$-$293~keV.  The pulsations are most prominent between 50$-$102~keV (panel
c, starting at $\sim$130~s). The bursts have not been removed here from the data. The dashed lines in panels (a) to (c) indicate the times of the pulse maxima
(as calculated using barycentered time).}
\label{fig:lc_pulse}
\end{figure}

During our search for untriggered events in the first trigger interval
of 2009 January 22 from \sgr, we found strong apparent periodic
modulations in the enhanced emission period from \trig+130 to 160 s in
the 50$-$102 keV data of detector NaI\,0 (see panel (c) of
Figure~\ref{fig:lc_pulse}). This is the first time to our knowledge
that pulsations {\it unrelated} to a giant flare from a magnetar were
clearly seen in the persistent emission of an SGR, in energies up to 100~keV.
To search for a coherent pulse period, we performed a timing analysis
over the entire enhancement interval.  We first eliminated the times
of all the bursts found via our untriggered burst search and converted
the remaining burst-free intervals to the solar system barycenter. For
each burst we removed 1~s centered at the burst peak; this elimination
resulted in a ``loss'' of $\sim$21~s during the interval
\trig+90$-$220~s. As the majority of the SGR bursts had durations
$<100$ ms, our method removed any effect of the burst contributions in
the time series. We then generated a Lomb-Scargle periodogram
\citep{lom75,sca82} over a range of periods from 0.1~s to 10~s using
CTIME Burst data in the 50$-$102~keV band. We found a very significant
signal with a Lomb power of 72.6 (chance occurrence probability, P$_c
\simeq10^{-16}$) at a period of 2.0699~$\pm$~0.0024~s, which is
consistent with the spin period of \sgr. Further, to confirm our
detection, we also employed the Z$^2$$_m$ test \citep[with m =
2;][]{buc83} on the burst-free and barycentered TTE data. We find a
coherent signal (with a Z$^2$$_{m=2}$ power of 266, P$_c
\simeq10^{-23}$) at the same period. Our spin period measurement is
consistent with the one found for \sgr~using contemporaneous X-ray
data \citep[{\it Swift}/XRT,][]{kui09, isr09} and radio data
\citep[obtained with Parkes;][]{bur09}.  Therefore, we clearly confirm
with the detection of these hard X-ray pulsations that the enhanced
persistent emission seen in the inset of Figure~\ref{fig:lc_ch14}
originates from \sgr.

Next we searched in the enhanced persistent emission for evolution in
the intensity of the pulsations using a sliding boxcar technique. We
found that the pulsed signal peaks over a 90~s interval, from
\trig$+120$ to $210$~s, which encompasses the peak of the enhancement.

Finally, we searched for any other intervals exhibiting pulsed
emission in the burst-free continuous CTIME data of 2009 January 22
and during the four subsequent days, using a sliding boxcar of 120 s
with 10 s steps. We did not find any additional
statistically-significant pulsed emission.
For the entire search and for all the timing analysis reported here,
we used more precise spin ephemeris obtained by contemporaneous
{\it Swift}/XRT, Chandra, XMM-Newton and Suzaku observations
(G.L. Israel et al. 2010, in preparation).

\subsection{Pulse Profiles}\label{sec:pprofile}

To investigate the evolution of the pulse profiles with energy, we
folded the burst-free TTE data spanning 120~s (from \trig+100~s to
\trig+220~s, which includes the strongest pulsation period as found above) with the 
spin ephemeris of \sgr.  We estimated the background level using the data 
segment between \trig~to \trig+60~s. Figure~\ref{fig:pptte} shows the 
source pulse profiles during the enhanced emission interval in six energy bands 
that have the same logarithmic width. The pulse profiles above 110~keV are
consistent with random fluctuations, and thus not shown.

Figure \ref{fig:pptte} indicates that the \sgr~pulse profiles in the
three lowest energy bands are most likely complex (multi-peaked).
While the two lowest energy band profiles are dominated by the
structure around phase 0.7-0.8 (indicated by the dotted lines in
Figure \ref{fig:pptte}), in the 14$-$22~keV band we see the emergence
of another structure around phase 0.0 (indicated by the dashed lines
in Figure \ref{fig:pptte}). This pulse becomes equally prominent in
the 22$-$33 keV range and then dominates in the 33$-$50 keV band. The
pulse profile changes remarkably in the 50-74 keV band, which is the
most statistically significant of all the energy bands investigated,
and is distinguished by a broad structure that peaks at around phase
0.0. The 74$-$110~keV profile resembles the 50-74 keV one.  As noted
above, the pulse profile above 110~keV is consistent with random
fluctuations.  Therefore, {\it our results set an observed upper
energy bound of 110~keV for the hard X-ray pulsations in \sgr} during
this enhanced emission episode.

\subsection{Pulsed Fraction}

We computed the RMS pulsed fraction using a Fourier based approach as
described in \citet{woo07}. In summary, we take the Fourier transform
of each pulse profile, then we calculate the RMS pulsed flux by taking
the Fourier coefficients of up to third harmonic into account, and
finally obtain the pulsed fraction values by dividing the RMS pulsed
flux by the phase-averaged flux. In Figure~\ref{fig:pfall}, we show
the pulsed fraction spectrum of \sgr~in the same energy bands as in
Figure \ref{fig:pptte}.

Although marginally significant, there is an indication of a minimum
in the RMS pulsed fraction around $\sim$30~keV. The RMS reaches its
maximum value of 0.55 $\pm$ 0.12 in the 50$-$74 keV band, and then
dips below detection at energies greater than $\sim$110~keV.  We will
discuss the implications of these results along with the results of
our spectral analysis in Section \S\ref{sec:sum}.

\subsection{Search for High-Frequency QPOs}

We also searched the period of enhanced emission for any signs of
high-frequency quasi-periodic oscillations (QPOs) similar to those
seen in the aftermath of SGR giant flares and attributed to the
excitation of global seismic modes \citep{isr05, str05, str06,
wat06}. Using TTE data from the three detectors (NaI\,0, 1 and 3) with
smallest detector zenith angles to the source (and not occulted by the
LAT), we selected photons with energies less than 100~keV, where the
enhanced emission dominates.  We searched the burst-free data set for
periodic and quasi-periodic oscillations (1~Hz and 2~Hz resolution)
and found no significant signals even on timescales as short as 1~s.
During the period when the emission is strongest (\trig+100 to 200~s),
the 3$\sigma$ upper limit on the amplitude of QPOs with frequencies in
the range 100$-$4096~Hz is 7.5\% RMS.

The upper limits are less constraining for frequencies below 100~Hz. A
rotational phase dependent search also revealed no significant
signals.  Finally, with the bursts included, we searched for
short-lived QPOs excited by each event: again, we found nothing
significant.

%%%%%%%%%%%%%%%%%%%%%%%%%%%%%%%%%%%%%%%%%%%%%%%%%%%%%%%%%%%%%%%%%%%%%%%%%%%%%%%
%%%%%%%%%%%%%%%%%%%%%%%%%%%%%%%%%% SECTION 3 %%%%%%%%%%%%%%%%%%%%%%%%%%%%%%%%%% 
%%%%%%%%%%%%%%%%%%%%%%%%%%%%%%%%%%%%%%%%%%%%%%%%%%%%%%%%%%%%%%%%%%%%%%%%%%%%%%%
\section{Spectral Properties of Pulsed Hard X-rays}\label{sec:spec}
\subsection{Time-Integrated and Time-Resolved Spectral Analysis}

We analyzed time-integrated and time-resolved spectra of the enhanced
emission, using the {\it RMFIT (3.1rc1)} spectral analysis software
developed for the GBM data analysis\footnotemark{}.
~\footnotetext{R.S.~Mallozzi, R.D.~Preece, \& M.S.~Briggs, "RMFIT, A
Lightcurve and Spectral Analysis Tool," \copyright 2008
Robert~D.~Preece, University of Alabama in Huntsville, 2008} Similar
to the timing analysis, we excluded all bursts identified with the
untriggered search within the enhancement period. Here, we removed up
to 3 seconds of data per burst centered at the burst peak, to account
for spectral contributions from the wings of each burst; this
elimination resulted in a loss of $\sim$25~s during the interval
\trig+70$-$220~s. We note that although some weak bursts may still be
included in our enhanced emission spectra, the very small intensities of these
bursts have practically no effect on our spectral
analysis results.  The spectral study of the untriggered bursts within
the enhanced emission indicates that their spectra are different from
the enhanced emission spectra (A.J. van der Horst~et~al. 2010, in
preparation).  For this analysis, we used only the CSPEC Burst data
(8.6$-$897~keV) of detector NaI\,0, which initially had the smallest
detector zenith angle to the source (15\degrees) and to which the
source was visible through most of the enhanced emission.

Since the Detector Response Matrices (DRMs) of GBM are time dependent
due to the continuous slewing of the spacecraft, a DRM should be
generated for every 2-3 degrees of slewing (corresponding to every
$\sim$20$-$50~s of data). For this analysis, we generated DRMs for
every 50~s starting from \trig, using GBMRSP {\it v1.7}
\citep[see][for a detailed description of the GBM response
generation]{mee09}.  We used a DRM generated at $T_0 + 150$~s for the
time-integrated spectrum (72$-$248~s), and three DRMs generated at
$T_0 + 100$, 150, and 200~s, respectively, for the time-resolved
spectra: each DRM was centered at the mid-time of the accumulation
time span of each spectrum.  The background spectrum was determined by
fitting a third-order polynomial function to each energy channel using
the burst-free intervals (\trig$-286$ to \trig$-43$~s, 1008$-$1196~s,
and 1941$-$2427~s), with a total accumulation time of 896~s.

We found clear evidence for spectral curvature below 100~keV in the
time-integrated spectrum of the entire burst-free enhancement period
(72$-$248~s): a single power law thus resulted in a very poor fit. We
employed five other spectral models; cut-off power law, power law +
blackbody, optically-thin thermal bremsstrahlung, and single/double
blackbody. We found that the time-integrated spectrum is best
described by a power law + blackbody (see Figure \ref{fig:nufnu}). All
other spectral models did not provide better fits mainly because they
failed to fit the lower energy excess $\lesssim10$~keV. The best-fit
spectral parameters of a power law with an additional blackbody are
shown in Table~\ref{tab:spec_result}. Adding a blackbody (with $kT
=18~\pm$ 4~keV) to a power law resulted in the most significant
improvement in Cash statistics \citep{cas79} over a single power law
($\Delta$C-stat = 13.5 for 2 degrees of freedom, corresponding to an
improvement of 3.25$\sigma$).

The average energy flux over the entire enhancement is $(6.5\pm2.4)
\times 10^{-8}$~erg~cm$^{-2}$~s$^{-1}$ (in 8$-$150~keV), of which the
blackbody component accounts for 19\%.  As stated earlier, the
distance estimate to the source is not well constrained; however,
assuming a source distance of $\sim5$~kpc, we estimate a total
isotropic emitted energy of $2.9 \times 10^{40}D_5^{~2}$~erg for the
entire persistent emission (8$-$150~keV) during the enhancement.

To investigate the evolution of the blackbody component and of the
source's spectral properties in general, we divided the enhanced
emission period into three time intervals of $\sim$50~s each:
74$-$117~s, 122$-$169~s, and 173$-$223~s after the trigger time. The
stopping time of the last spectrum was \trig+223~s, because the source
was occulted by the LAT for NaI\,0 around \trig+225~s.  We employed
the same set of photon models as the time-integrated analysis
described above.  The first spectrum was best fit by a single power
law with no evidence of a blackbody or any curvature.  The second and
third spectra, on the other hand, were best described by power law +
blackbody models.  In the second spectrum (the peak of the
enhancement) the additional blackbody component was statistically most
significant (see Figure~\ref{fig:nufnu}), and remained significant in
the third spectrum as well.  The ratio of the blackbody flux to the
total flux (8$-$150~keV) was found to be 34\% in both intervals. The
indices of the underlying power-law component, and the blackbody
temperature also remained constant, at $\sim$$-2.1$ and $\sim$17~keV,
respectively (within uncertainties; see also Table 1), while the
power-law amplitude tracked the photon flux.
					
We note that the last two time-resolved spectra were also fitted by a
cut-off power law model with the fit being statistically as good as
the power law + blackbody model. The difference between the two models
becomes apparent only at energies $\gtrsim$200~keV, where the count
rates of our data drop dramatically.  For a further comparison, we
simulated spectra with the best-fit cut-off power law model for the
second spectrum, folded them through the GBM NaI\,0 DRM, and fitted
the simulated spectra with a power law + blackbody (and vice versa).
We found no indication of significant statistical preference between
these two models due to the low count rates at higher energies.  The
best-fit cut-off power law parameters for the second and third spectra
are also shown in Table~\ref{tab:spec_result}.

Finally, we also analyzed the spectrum integrated over the second and
third time intervals (\trig+122 to 223~s), in which the blackbody
component was found to be very significant.  The blackbody + power law
model parameters for this combined spectrum are listed in
Table~\ref{tab:spec_result}.  The results were consistent with the
time-resolved analysis of the individual spectra described above, and the
statistical significance of the blackbody component was similar to
that of the second spectrum ($\Delta$C-stat = 42.5, corresponding to a
6.2$\sigma$ improvement).  A cut-off power law model also provided an
adequate fit for this combined spectrum; the best-fit parameters
of the cut-off power law are also listed in
Table~\ref{tab:spec_result}.

\subsection{Phase-Resolved Spectral Analysis}

We performed spin-phase-resolved spectral analysis of the pulsed
enhanced emission, as follows: we co-added the burst-free spectrum of
each pulse (in \trig+122 to 223~s, corresponding to second and third
time-resolved spectra) using TTE data and extracted a phase-maximum
spectrum (between phases 0.75$-$1.25 in Figure \ref{fig:pptte}) and a
phase-minimum spectrum (between phases 0.25$-$0.75). The spin phase
for each photon was calculated using barycentered times, as was done
for the timing analysis. We calculated the background spectrum, from
the burst-free interval at \trig~to \trig+60~s.

The spectra of both the phase minimum and maximum were adequately
fitted with power law + blackbody models, where we kept the power-law
indices and the blackbody temperatures linked.  The values of the
linked parameters found in the fit were consistent (within 1$\sigma$)
with those of the time-integrated spectra (see
Table~\ref{tab:spec_result}, 122$-$233~s). However, the blackbody
component was more significant in the phase-maximum spectrum than in
the phase-minimum spectrum. The contributions of the blackbody flux to
the total flux were (52$\pm$18)\% and (35$\pm$18)\%, in the
phase-maximum and phase-minimum spectra, respectively.

%%%%%%%%%%%%%%%%%%%%%%%%%%%%%%%%%%%%%%%%%%%%%%%%%%%%%%%%%%%%%%%%%%%%%%%%%%%%%%%
%%%%%%%%%%%%%%%%%%%%%%%%%%%%%%%%%% SECTION 5 %%%%%%%%%%%%%%%%%%%%%%%%%%%%%%%%%% 
%%%%%%%%%%%%%%%%%%%%%%%%%%%%%%%%%%%%%%%%%%%%%%%%%%%%%%%%%%%%%%%%%%%%%%%%%%%%%%%
\section{Summary and Discussion}\label{sec:sum}

We report here the discovery of coherent pulsations in the persistent
hard X-ray emission from \sgr~in the {\it Fermi}/GBM data lasting
$\sim$150~s. Coherent pulsations with a 55\% RMS pulse fraction have
never been detected in the persistent emission at these high energies
from a magnetar as yet. These pulsations were detected {\it only} at
the onset of a major bursting episode and were not directly related to
a major burst or flare from the source. The pulse period is consistent
with the spin period of \sgr~as measured with contemporaneous {\it
Swift}/XRT observations, thus confirming
\sgr~as the origin of the enhanced emission. We estimate the total
isotropic-equivalent emitted energy during the persistent emission
(i.e., excluding burst contributions) to be $2.9 \times 10^{40}D_5^{~2}$~erg.
The thermal component accounts for 19\% of
the total emitted energy; i.e., $5.6 \times 10^{39}D_5^{~2}$~erg is emitted as
blackbody.

The fact that this enhanced emission was detected at the onset of a
major bursting episode without evidence of direct association to any
particular burst or flare immediately before the emission is very
intriguing. Intermediate flares with pulsating tails were observed
from SGR\,1900+14 \citep{ibr01,len03} and very recently from
\sgr~\citep[$\sim$6~hours after \trig]{mer09}.  Thermal components
were also found in the decaying tails of intermediate events from
SGR\,1900+14 with much lower blackbody temperatures of $\sim$2~keV
\citep{len03}.  The thermal component of the enhanced emission we
report here is hotter (17~keV), exhibits a strong dependence of the
pulse profile with energy with a very high RMS pulsed fraction (up to
55\%), and is clearly not associated with a decaying event tail.  {\it
Energetically, however, the fluence of this enhanced emission is
comparable to that of tail emission of the intermediate flares from
SGR\,1900+14.}

\begin{figure}[htbp]
%\epsscale{0.9}
\begin{center}
\plotone{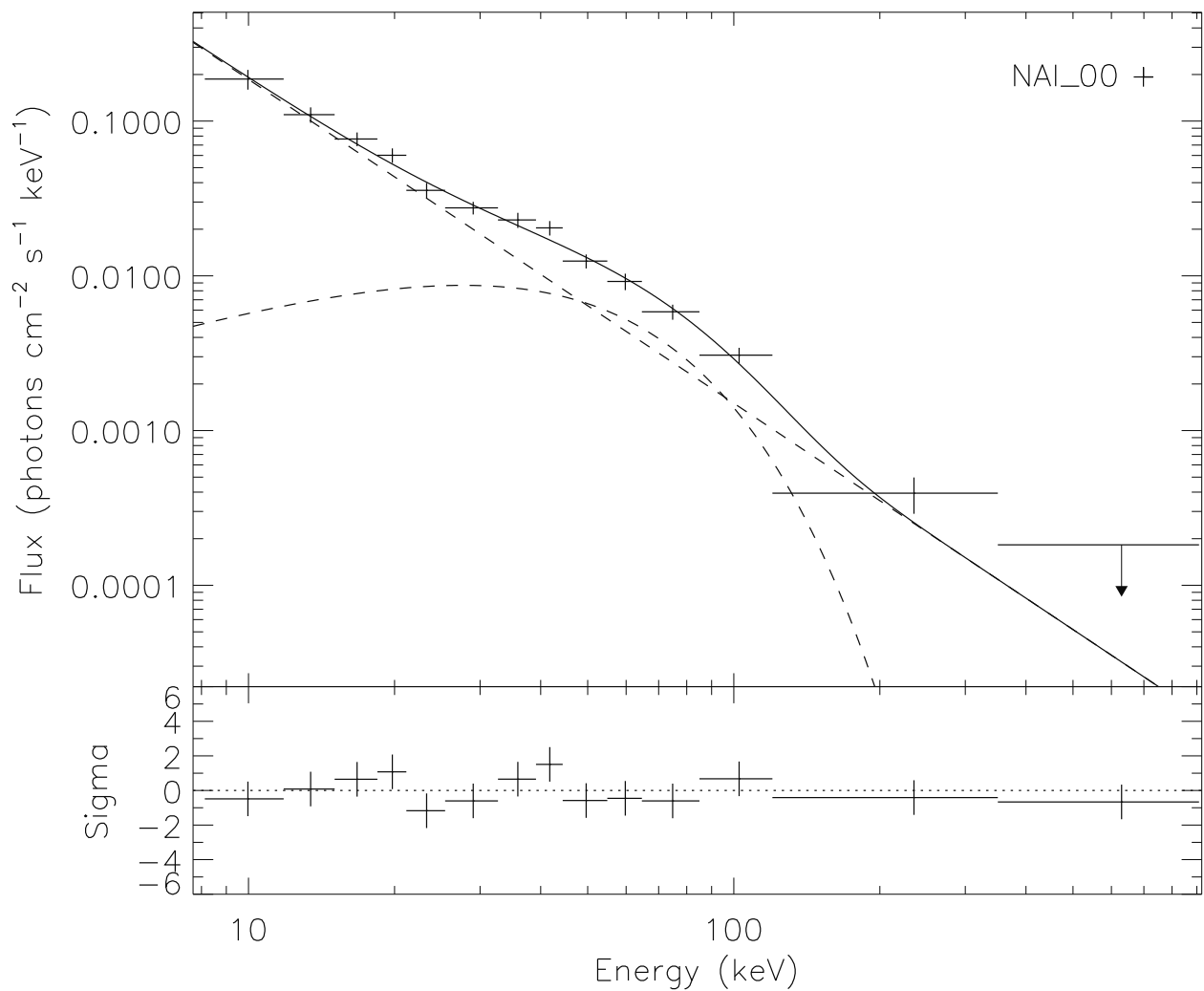}
\caption{The photon spectrum of the time interval \trig+122 to 169~s.  The blackbody and power-law components are shown separately with dashed curves.
The data are binned for display purpose only. A 3$\sigma$ upper limit is shown for the highest energy bin.}
\label{fig:nufnu}
\end{center}
\end{figure}

Our timing analysis showed that the detection of pulsations is most
significant in the 120$-$210~s interval after trigger. We find that
the spectrum requires a blackbody component along with a power law
between 122$-$223~s, which is consistent with the time interval of the
most significant detection of pulsations. Moreover, as determined by
the energy dependent pulse profiles and RMS pulsed fractions, we find
that the high-energy pulsations are most significant in the
50$-$74~keV range.  Strikingly, the blackbody component of the
enhanced persistent emission spectrum peaks at around 51~keV (i.e.,
the Wien peak of 17~keV, see Figure~\ref{fig:nufnu}). These two
independent pieces of evidence lend strong support for a blackbody
radiation component to account for the curvature in the spectrum of
the enhanced emission.

In our spin-phase-resolved spectral analysis, we find that the
blackbody flux to the total emission is (52$\pm$18)\% and
(35$\pm$18)\% in the phase-maximum and phase-minimum spectra,
respectively. This also suggests that a major contribution to the
observed pulsations is from the blackbody component. If we assume a
surface hot-spot during this pulsating interval, then the best-fit
blackbody corresponds to an effective radiating area (as projected on
the plane of the sky, far from the star) of $S_\infty =\pi D^2
F/(\sigma T^4) \approx 0.046D_5^{~2}\;{\rm km}^2$, where $T \approx
2\times 10^8\;^\circ$K ($kT \approx 17\;$keV) is the observed
(gravitationally redshifted) temperature. We have used here the
blackbody flux at the peak of the pulsations (i.e., phase maximum; $F
\approx 5\times 10^{-8}\;{\rm erg\;cm^{-2}\;s^{-1}}$) where the
hot-spot is expected to be relatively close to face-on, in order to
minimize the effects of projection and gravitational lensing by the
neutron star, so that $S_\infty$ would be relatively close to the
physical area, $S$, of the hot spot on the neutron star surface. For a
circular hot-spot, this corresponds to a radius of
$\sim$120$D_5$\,m.

The rotational energy of magnetars is insufficient for powering their
observed emission, since they all have long rotation periods, and
their spin-down luminosity is much lower than their observed
luminosity. Owing to their slow rotation, only a very small fraction
($\sim R_{\rm NS} \Omega /c\sim 10^{-5}-10^{-4}$) of the magnetic flux
threading the neutron star corresponds to open field
lines~\citep{bel07}. Since the internal field of magnetars can be
significantly stronger and more tangled (or twisted) than the external
(largely dipole) field, the transfer of magnetic helicity from the
interior to the exterior of the neutron star powers magnetar
activity~\citep{bel07}. As the internal field twists the stellar
crust, the magnetosphere also becomes twisted, possibly in a
complex manner~\citep{tho02}.
 
~\citet{bel07} have shown that the rate of energy dissipation in the
twisted magnetosphere is $L_{\rm d}= I \Phi_{e}\sim 10^{38}\Delta \psi
(B/10^{15}\,{\rm G}) ( a/R_{\rm NS})^{2}(e\Phi_{e}/{\rm
10\,GeV})\;{\rm erg\;s^{-1}}$, where $I$ is the net current through
the corona, $\Phi_{e}$ is the voltage along the twisted magnetic
lines, $a$ is the size of a twisted region on the stellar surface and
$\Delta \psi$ characterizes the strength of the twist.  Identifying
$a$ with the inferred size of the hot-spot ($\sim 120D_5\,$m)
would imply $a/R_{\rm NS} \sim 10^{-2}$ which is inconsistent with the
observed luminosity of the spot, $L_{\rm d} \sim 10^{38}D_5^{~2}\;{\rm
erg\;s^{-1}}$ since $e\Phi_{e} \lesssim 10\;$GeV is expected (limited
by pair creation) and $\Delta\psi \lesssim 1$ is required for global
stability.

It may be possible for the magnetic twist to grow to a global
instability level during a highly active bursting period due to
frequent starquakes \citep[i.e., $\Delta \psi\gtrsim 1$;][]{bel07}.
As the magnetosphere untwists, a large amount of energy must be
dissipated \citep{lyutikov}.  A small ``trapped fireball'' -- plasma
of $e^\pm$ pairs and photons confined by a closed magnetic field
region -- could then potentially account for the inferred hot-spot,
and in particular its roughly constant temperature and size. Confining
a ``fireball" of energy at least comparable to that emitted by the
observed blackbody component, $E_{\rm iso,BB} \approx 5.6 \times
10^{39}D_5^{~2}\;$erg, within a region of radius $a \sim 120D_5\;$m
requires $E_B(a) = \frac{1}{6}a^3B^2 > E_{\rm iso,BB}$ or $B \gtrsim
1.4 \times 10^{14}(a/120D_5\,{\rm m})^{-3/2}(E_{\rm iso,BB}/
5.6\times 10^{39}D_5^{~2}\,{\rm erg})^{1/2}\;{\rm G} \approx 1.4 \times
10^{14}D_5^{-1/2}\;$G. This is consistent with the surface dipole
field of $B\approx 2.2\times 10^{14}\;$G inferred from the measured
$P\dot{P}$ \citep{cam07}. Therefore, a sufficiently small closed
magnetic loop anchored by the crust could provide the required
confinement.

Although the neutron star surface is relatively cold, hot spots may
naturally form on the stellar surface since the energy dissipated in
the corona is thermalized as it passes through the denser atmosphere
and reaches the stellar crust.
The large pulsed fraction implies that the emitting region responsible
for the pulsations, which we identify with one or two hot-spots on the
stellar surface, is mostly or totally obscured during certain rotation
phases. This implies that if we are observing emission from two
hot-spots (of similar temperature) then they cannot be located too far
from each other (for example, they cannot be antipodal), as most of
the stellar surface is visible to an observer at infinity at any given
time because of strong gravitational lensing by the neutron star.

\begin{figure}[htbp]
%\epsscale{0.8}
\centerline{
\plotone{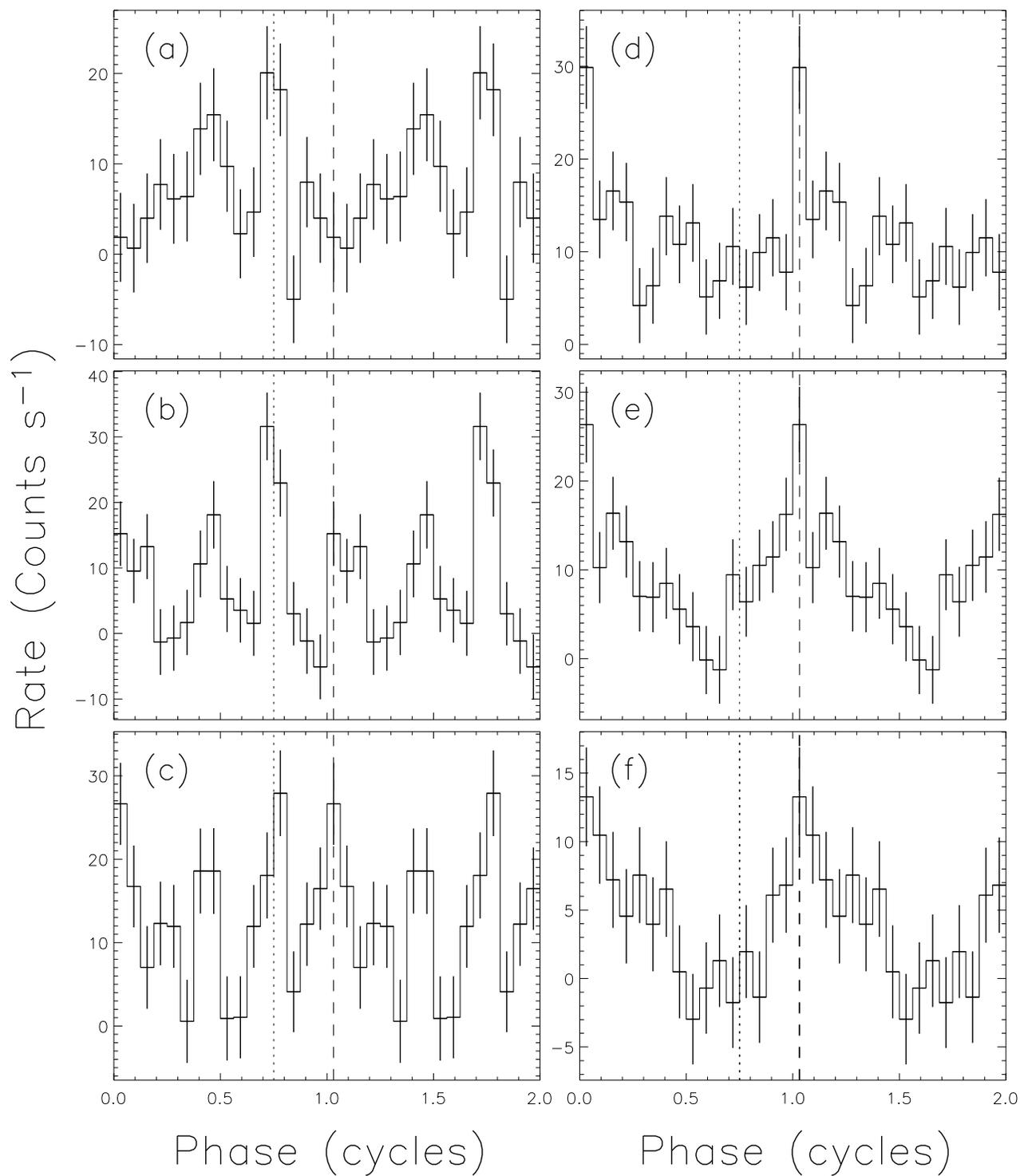}}
\caption{Pulse profiles of \sgr~in equal logarithmic energy intervals;
(a) 10$-$14~keV, (b) 14$-$22~keV, (c) 22$-$33~keV, (d) 33$-$50~keV,
(e) 50$-$74~keV, and (f) 74$-$110~keV.  Two cycles are plotted
for clarity. The vertical dotted and dashed lines are explained in \S\ref{sec:pprofile}.
\label{fig:pptte}}
\end{figure}

\begin{figure}[htbp]
%\epsscale{0.9}
\centerline{
\plotone{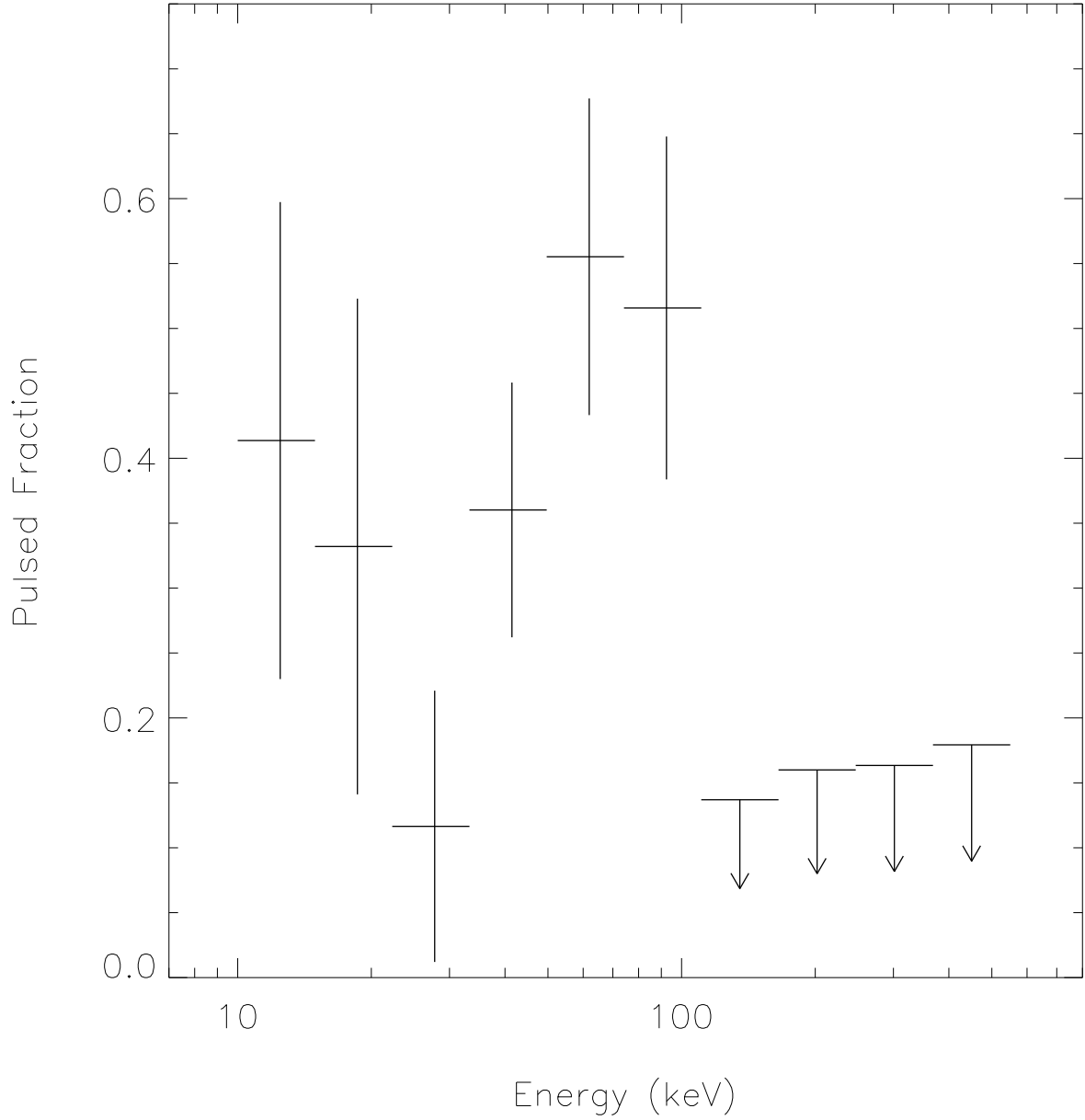}}
\caption{Evolution of RMS pulsed fraction of \sgr~as a function of energy.  Uncertainties are 1$\sigma$.
The energy bands are the same as those used in Figure~\ref{fig:pptte}.
\label{fig:pfall}}
\end{figure}

Moreover, we might be observing two hot-spots, possibly corresponding
to the two footpoints of a twisted magnetic flux tube (or two
magnetically confined regions), where the second hot-spot is somewhat
cooler and dominates the pulsed emission below $\sim$20\,keV. This
might explain the energy dependence of the pulse profiles (see
Fig.~\ref{fig:pptte}), with the appearance of a second peak at lower
energies (below $\sim$33\,keV, at a phase of $\sim$0.75), as well as
the increase in the RMS pulsed fraction at the lowest energies (with a
local minimum around $\sim$28\,keV; see Fig.~\ref{fig:pfall}) and the
peak at $\sim$60\,keV corresponding to the hotter hot-spot.

This blackbody emission is expected to be accompanied by non-thermal,
high-energy radiation produced by collisionless
dissipation. \citet{bel07} estimate that the luminosities in the
high-energy and blackbody components should be comparable. This is in
good agreement with our observations of \sgr, where the high-energy
(power-law) and blackbody contribution to the total luminosity were
found to be 65\% and 35\%, respectively for the time-resolved
spectra. These contributions were 48\% and 52\% in the phase-resolved
pulse-maximum spectrum. 

In conclusion, the area of the blackbody emitting region is the smallest ``hot spot" measured for a magnetar, which likely arises from magnetically confined hot plasma on the neutron star surface, possibly caused by the gradual dissipative untwisting of the magnetosphere \citep{lyutikov}. If the total radiated energy was initially confined to the inferred extremely small size of the enhanced emission region (as in a mini ``trapped fireball'' scenario), this would indicate a very
large magnetic energy density (for $B \gtrsim 1.4 \times 10^{14}D_5^{-1/2}\;$G), similar to the ``trapped fireball'' model for the tails of SGR giant flares. 
The observed enhanced emission that we report here is much less energetic than a giant flare tail, while its energy is comparable to the tail energy of intermediate events and at the high end of typical SGR bursts.  
Despite some distinct properties, the enhanced emission of \sgr~carries various flavors of all three SGR phenomena, and thus it is most likely related to the very pronounced bursting activity that immediately followed it. \\

%Acknowledgment
We would like to thank the referee, Dr. F.~Camilo, for his valuable comments.
 This publication is part of the GBM/Magnetar Key Project
(NASA grant NNH07ZDA001-GLAST, PI: C. Kouveliotou). We thank
G.L. Israel and A. Tiengo for providing the precise spin ephemeris and
source distance, respectively, prior to their publication. YK and EG
acknowledge EU FP6 Transfer of Knowledge Project ``Astrophysics of
Neutron Stars'' (MTKD-CT-2006-042722). JG gratefully acknowledges a
Royal Society Wolfson Research Merit Award. ER-R thanks the Packard
Foundation for support.  AJvdH was supported by an appointment to the
NASA Postdoctoral Program at the MSFC, administered by Oak Ridge
Associated Universities through a contract with NASA.

%%%%%%%%%%%%%%%%%%%%%%%%%%%%%%%%%%%%%%%%%%%%%%%%%%%%%%%%%%%%%%%%%%%%%%%%%%%%%%%
%%%%%%%%%%%%%%%%%%%%%%%%%%%%%%%%%% Reference %%%%%%%%%%%%%%%%%%%%%%%%%%%%%%%%%% 
%%%%%%%%%%%%%%%%%%%%%%%%%%%%%%%%%%%%%%%%%%%%%%%%%%%%%%%%%%%%%%%%%%%%%%%%%%%%%%%

\newpage
%%%%%%%%%%%%%%%%%%%%%%%%%%%%%%%%%%%%%%%%%%%%%%%%%%%%%%%%%%%%%%%%%%%%%%%%%%%%%
%%%%%%%%%%%%%%%%%%%% Figures %%%%%%%%%%%%%%%%%%%%%%%%%%%%%%%%%%%%%%%%%%%%%%%%
%%%%%%%%%%%%%%%%%%%%%%%%%%%%%%%%%%%%%%%%%%%%%%%%%%%%%%%%%%%%%%%%%%%%%%%%%%%%%

\newpage
%%%%%%%%%%%%%%%%%%%%%%%%%%%%%%%%%%%%%%%%%%%%%%%%%%%%%%%%%%%%%%%%%%%%%%%%%%%%%
%%%%%%%%%%%%%%%%%%%% Tables %%%%%%%%%%%%%%%%%%%%%%%%%%%%%%%%%%%%%%%%%%%%%%%%%
%%%%%%%%%%%%%%%%%%%%%%%%%%%%%%%%%%%%%%%%%%%%%%%%%%%%%%%%%%%%%%%%%%%%%%%%%%%%%

\begin{table}[htbp]
\caption{\small Spectral parameters of the enhanced persistent emission period of \sgr.}
\tabcolsep=2pt
\scriptsize
\begin{center}
\vspace{-2ex}
\begin{tabular}{cccccccccccc}
\hline \hline \\[-2ex]
Time & \multicolumn{2}{c}{Power Law (PWRL)$^{^1}$}   && \multicolumn{2}{c}{Blackbody (BB)} && \multicolumn{2}{c}{Energy Flux$^{^3}$} &   & \multicolumn{2}{c}{Cut-off Power Law$^{^2}$} \\
since \trig& &  && \multicolumn{2}{c}{} && \multicolumn{2}{c}{\tiny (10$^{-8}$ ergs/cm$^{2}$-s)} &  &
&    \\
\cline{2-3}\cline{5-6}\cline{8-9}\cline{11-12} \\[-2ex]
& 	$A$	&	$\gamma$ && $N$ & $kT$ & $\Delta$C-stat	 & PWRL & BB & F$_{\rm BB}$/F$_{\rm total}$ & $\alpha$ & E$_{\rm peak}$\\
 s & \tiny ($\times$10$^{-4}$~ph/s-cm$^2$-keV)	&&& \tiny	($\times$10$^{-5}$~ph/s-cm$^2$-keV) & \tiny (keV) & & & & & &\tiny (keV)  \\ \\[-2ex]
\hline 
 72$-$248 & 	10.53 (1.96) & 	$-$2.06 (0.10) && 	1.23 (0.96) & 	17.7 (3.8) & 	13.5	 & 5.30 (2.37)	 & 1.22 (0.28) & 0.19 (0.08) & $-$ & $-$ \\
\hline									
74$-$117 & 	5.20 (1.30) &	$-$2.15 (0.17)&&	No BB	& $-$ & $-$	&	2.85 (3.30)	& $-$ & $-$	&  $-$ & $-$	\\
122$-$169	 & 15.02 (3.05)	 & $-$2.09 (0.11)	 && 4.45 (1.58)	 & 17.4 (1.7)	 & 42.9	 & 7.82 (4.47) & 4.08 (0.65) & 	0.34 (0.14)   & $-$1.30 (0.14)& 68 ( 7)\\
173$-$223	 & 9.18 (3.28)	 & $-$2.14 (0.19) & &	3.49 (2.12)	 & 16.4 (2.7) & 	15.3 & 5.05 (3.75)	 & 2.59 (0.72) & 0.34 (0.19) & $-$1.33 (0.25)& 59 (10) \\
\hline									
122$-$223	 & 13.27 (2.29)	 & $-$2.08 (0.10)	 && 3.74 (1.50)	 & 16.5 (1.8)	 & 42.5 & 	6.81 (2.99) & 2.84 (0.46) &  0.29 (0.10) & $-$1.41 (0.13)& 65 ( 7)    \\
 \hline
\end{tabular}
\begin{minipage}{0.98\textwidth}
\vspace{6pt}
\small{$^1$Power Law Model: $f$(E)$ = A$(E/100\,keV)$^\gamma$} \\
\small{$^2$Cut-off Power Law Model: $f$(E)$ = A$\,exp[$-$E($2 + \alpha$)/E$_{\rm peak}$](E/100\,keV)$^\alpha$}\\
\small{$^3$ Flux is calculated in 8$-$150~keV.}\\
\small{$^4$ 1-$\sigma$ uncertainties are shown in parentheses.  $\Delta$C-stat shows an improvement in Cash statistics by adding a blackbody with 2 degrees of freedom to a power law.  Cut-off power law parameters are shown for the cases where the model also provided adequate fits.}
\end{minipage}
\end{center}
\label{tab:spec_result}
\end{table}

\end{document}